\documentstyle[preprint,aps]{revtex}
\begin{document}

\def\la{\mathrel{\mathpalette\fun <}}
\def\ga{\mathrel{\mathpalette\fun >}}
\def\fun#1#2{\lower3.6pt\vbox{\baselineskip0pt\lineskip.9pt
\def\beq{\begin{equation}}
\def\eeq{\end{equation}}
\def\beqa{\begin{eqnarray}}
\def\eeqa{\end{eqnarray}}
\def\tr{{\rm tr}}
\def\x{{\bf x}}
\def\p{{\bf p}}
\def\k{{\bf k}}
\def\z{{\bf z}}
         
\ialign{$\mathsurround=0pt#1\hfill##\hfil$\crcr#2\crcr\sim\crcr}}}

\preprint{\vbox{
\hbox{CERN-TH/97-123}
\hbox{CWRU-P5-1997}
}}

\title{Hybrid Inflation and Particle Physics}

\author{Gia Dvali$^a$, Lawrence M. Krauss$^{a,b,}$\footnote{
also Department of Astronomy, CWRU} and Hong Liu$^{a,b}$\footnote{Address
after October 1st: Blackett Laboratory, Imperial College,
London, SW7 2BZ, U.K.}} 

\address{$^a$Theory Division, CERN, 1211 CH Geneva 23, Switzerland}

\address{$^b$Physics Department,
Case Western Reserve University,
Cleveland OH 44106-7079\footnote{Permanent address for LMK} 
}

\date{\today}

\maketitle

\begin{abstract}

The prototype hybrid SUSY SU(5) inflation models, while well motivated 
from particle physics, and while allowing 
an acceptable inflationary phase with little or no fine tuning, are shown 
to have two fundamental phenomenological problems. (1) They inevitably 
result in the wrong vacuum after inflation is over; and (2) they do 
not solve the monopole problem.  In order to get around the first 
problem the level of complexity of these models  must 
be increased. One can also avoid the
second problem in this way.  We also demonstate another possibility
by proposing a new general mechanism 
to avoid the monopole problem with, or {\it without} inflation.

\end{abstract}


\newpage

Over the past 20 years, the interface between particle physics and 
cosmology has blossomed remarkably, spurred on by the 
development of Grand
Unified Theories (GUTs) in the 1970's.  These promised, for the first time, an
understanding of microphysics which would determine the equation of state for
the FRW expansion at very early times.  It was in this context that the
most influential modern theoretical
development in early universe cosmology occurred.  We
refer of course to inflation \cite{inflation}.

Nevertheless, in spite of this close connection between Guth's original
idea and the reigning particle theory of the day, many recent 
so-called particle physics inspired models for early universe cosmology are
often nothing of the sort.  Inflationary models are generally designed and
tuned not in response to particle physics issues, but rather cosmological 
ones. 
The constraints which govern the model parameters involve such issues
as the necessity for sufficient inflation, and the generation of subsequent
density perturbations. Often the motivation from 
known low energy physics questions is 
not immediately
apparent.

In this paper, we address what seems a very promising set of inflationary 
models
which have evolved from considerations of supersymmetry and supergravity, 
ideas which are central to current particle physics model building.  These
models, involving what has become known as hybrid inflation\cite{hybrid},
are based on a
generic characteristic of supersymmetric model building:
pseudo-flat directions in
scalar field potentials.  Moreover, they turn out to allow a novel 
graceful exit
from inflation which results in an acceptable scale for primordial density
perturbations without an apparent fine tuning of parameters, the bane of most
inflationary models. Finally, they have the very attractive feature that
the inflationary phase transition is related to the GUT transition, which is
again well motivated by particle physics.

Nevertheless, in spite of their origins in GUTs and supersymmetry, a number of
fundamental particle physics issues associated with hybrid inflation
have not received attention in the literature.   We
demonstrate that a consideration of such issues implies that
the  GUT prototypical hybrid inflationary models are generally not viable, and must
be supplemented by new complications.

Two generic problems face GUT models, and inflation was developed
in some sense in the effort to cure both of them.  First and foremost, GUT
symmetry breaking tends to inevitably produce stable magnetic monopoles, whose
abundance, in a cosmological context, is unacceptable.  Inflation was first
proposed in the context of resolving this problem.  Next, GUT models such as
$SU(5)$ can 
break in many different ways.  Since low-energy physics is described
by an $SU(3) \times SU(2) \times  U(1)$ model, it is clearly a potential 
phenomenological
disaster if $SU(5)$ breaks to $SU(4) \times U(1)$ 
preserving vacuum instead of the known
low energy vacuum configuration.  
Considerations of metastability of the ``false"
$SU(4) \times  U(1)$ vacuum in part led to the 
development of techniques which would
later be applied to inflationary models.

It may seem surprising, therefore, that simplest GUT prototype of hybrid
inflationary models, which
come closest perhaps to be inspired by current particle theory, in fact
generically do not resolve either of these issues.
If we preserve the attractve feature that the inflationary
transition is tied to the GUT transition, then, as we show here, the
prototype hybrid inflationary models are generically not phenomenologically
viable without either additional structures in the superpotential, or changing
the nature of the inflaton field.

First, let us review a minimal SUSY GUT version
of the canonical hybrid model\cite{hybrid}.  It is
given by a
superpotential, with two fields\cite{copeland},
the inflaton field, and the Higgs
field
in the present context to be
associated with GUT symmetry breaking
\cite{dss,lss}.
The simplest superpotential that leads to hybrid
inflation in the minimal supersymmetric $SU(5)$ is
\begin{equation}
W = {g\over 2}S{\rm Tr }\Sigma^2  - SM^2+ {h \over 3}{\rm Tr} \Sigma^3
\end{equation}
where $S$ is a gauge singlet field (the inflaton) and $\Sigma$ is an $SU(5)$
adjoint Higgs. From a particle physics perspective, the role of $S$ is to remove
the origin $\Sigma = 0$ from the vacuum manifold and break $SU(5)$. The 
role of the cubic invariant is to remove the unwanted 
continuous degeneracy and fix
the orientation of the $\Sigma$ VEV. 
In a globally supersymmetric limit the system admits the two 
degenerated minima with unbroken
$G_{3-2-1} = SU(3)\otimes SU(2)\otimes U(1)$
\begin{equation}
 \Sigma  = M\sqrt {{1 \over 15 g }} (2,2,2,-3,-3),~~        \label{2-3}
\end{equation}
and 
$G_{4-1}= SU(4)\otimes U(1)$ symmetries respectively
\begin{equation}
\Sigma = M\sqrt {{1 \over 10 g }} (1,1,1,1,-4).~~~\label{4-1}
\end{equation}
Supersymmetry breaking effects (e.g. in gravity-mediated scenarios)
remove the degeneracy, but the energy splitting is tiny (suppressed
by a factor $\sim ({M_{GUT} \over M_P})^2$ 
with respect to the potential barrier
$\sim M_{GUT}^4$).\footnote{As shown by Weinberg\cite{weinberg},
in the minimal $SU(5)$, provided the cosmological
constant in one of the vacua is fine-tuned to zero 
by adding an explicit constant
in the superpotential, the others in general appear to have negative
energy density. Since the energy splitting is small, the
gravitational effects can prevent 
the zero energy vacuum from  decaying\cite{cl}.}
 As a result tunneling is negligible and the minima
are stable for all practical purposes\cite{weinberg}. Which minima will be chosen by the
theory is thus determined in a cosmological context.

This system can lead to hybrid inflation for large values of the $S$
field. The scenario proceeds as follows. For large values  $S >> M/\sqrt{g}$, the
$\Sigma$ field becomes very heavy,  has a vanishing VEV and can be
integrated out. The effective superpotential for the remaining light 
$S$ field is just
linear
\begin{equation}
W_{inflation} = SM^2
\end{equation}
and the classical potential is exactly flat: giving a constant
vacuum energy density that breaks supersymmetry
\begin{equation}
V = M^4.
\end{equation}
The one-loop
corrected K\"ahler potential (for $ S >> M/\sqrt{g}$) is of the form
\begin{equation}
K = SS^+\left ( 1 - {g^2 3\over 2\pi^2}{\rm ln}{SS^+ \over m^2} \right  
)
\end{equation}
Since for large $S$, supersymmetry is broken, this simply translates  
into
the following effective potential for $S$\cite{dss}
\begin{equation}
 V(s) \simeq \Lambda^4 \left ( 1 + {g^2 3\over 2\pi^2}ln {SS^+ \over m^2}  
\right )
\end{equation}
which drives inflation\footnote{In a supergravity context, for the generic
K\"ahler metric, the above potential can receive non-trivial corrections
that may affect details of the inflationary dynamics\cite{copeland,gr,linderiotto} (slow-roll etc.).
Our conclusions, however,
are independent of the precise form of the inflaton potential, subject
to a general assumption that the GUT transition is triggered by $S$ at the end
of inflation.}
When $S$ drops below $S_c = M/\sqrt{g}$, the point $\Sigma = 0$ becomes
a local maximum. $\Sigma$ picks up a  
nonzero
VEV which grows to cancel the vacuum energy density. After this moment the system
rapidly relaxes to one of the minima and oscillates about it.
Note that the end of  
inflation
does not necessarily coincide with this transition but can happen much
earlier when the slow roll conditions break down.

We now demonstrate that the minimum
which the 
system chooses right after the phase transition is $G_{4-1}$ symmetric.
This conclusion is independent of the detailed structure of the
superpotential for $\Sigma$. 

 To see this let us consider the potential for $\Sigma$ in the background of the
$S$ field. We will be concerned by the structure of this potential at the moment
when $S$ drops to its critical value, after which $\Sigma$ gets
tachionic and begins to roll away from zero.
The effective potential for $\Sigma$ is
\begin{eqnarray}
V & = & h \sqrt{g}M {\rm Tr} \Sigma^*\Sigma^2 + h.c. + 
\left (g^2/4 - h^2/5 \right ) |{\rm Tr} \Sigma^2|^2 + 
h^2 {\rm Tr }\Sigma^{*2}\Sigma^2 
\nonumber 
\\ 
 &  &+ {g_{gauge}^2 \over 2} {\rm Tr} [\Sigma\Sigma^*]
\label{potsigma}
\end{eqnarray}
The last ($D$) term is automatically minimized for a diagonal $\Sigma$. For
the remaining terms only the first 
is phase-dependent. Thus 
irrespective of the sign of 
the parameters we can set $\Sigma$ real (the $S~ {\rm and}~ \Sigma$
phases will 
automatically adjust in such a way that to make the overall sign of 
the first term negative).
Then, for a fixed $S$, the potential is effectively a function of three
quantities: the absolute value $\sigma^2 =  Tr \Sigma^2$ and the two angles
(orbit parameters)\cite{orbit}
$\theta = {\rm Tr}\Sigma^3/\sigma^3$ and $\phi = {\rm Tr}\Sigma^4/\sigma^4$.
\begin{equation}
V = 2h \sqrt{g}M \sigma^3\theta + 
\left (g^2/4 - h^2/5 \right ) \sigma^4 + 
h^2 \sigma^4\phi \label{potorbit}
\end{equation}
The orientation of $\Sigma$ is entirely determined through
these orbit parameters, which take different values for the different
breaking patterns. It is well known \cite{minima} that the minima of (\ref
{potsigma}) (as well as of an arbitrary quartic or lower order 
gauge-invariant
polynomial of (real) $\sigma$) can only correspond to one of the maximal
unbroken subgroups: $G_{3-2-1}$ or $G_{4-1}$. The corresponding values of the
orbit parameters are $\theta_{3-2-1} = 1/\sqrt{30},~~~\phi_{3-2-1}=-7/30$
and  $\theta_{4-1} = 3/\sqrt{20},~~~\phi_{4-1}=-13/20$. As we have argued, the
sign of $\theta$ is irrelevant since the overall sign of the $\theta$-term will
be negative. Thus  at the moment when
$\sigma$ starts rolling away from zero, the cubic term dominates and 
the preferred 
orientation is the one that gives the largest value of $|\theta|$,
that is the $G_{4-1}$ direction. 
One may wonder whether this orientation can change for larger values 
of $\sigma$
 when the quartic term becomes significant. We can easily see that this is not
the case all way until the nearest minimum, which therefore corresponds to a
$G_{4-1}$ unbroken subgroup. First we note that the change from $G_{4-1}$ to
$G_{3-2-1}$ can not happen gradually, since this are the only possible minima
independently of the absolute value of $\sigma$. So 
$G_{4-1}$ will stay as the preferred direction until $\sigma$ grows to a
certain critical value $\sigma_c$  at which time the $\phi$ term will dominate
and
$G_{3-2-1}$ will become preferred. If $\sigma_c$ is larger than the value in
the closest minimum, $\sigma_0$,
$\sigma$ will be trapped before reaching $\sigma_c$. 
By the definition of $\sigma_c$,  $\sigma_0$ corresponds to
$G_{4-1}$ minimum. Thus, the system will automatically relax to the $G_{4-1}$
minimum, if $\sigma_c > \sigma_{4-1}$. For $S = S_c$ this is indeed the case:
\begin{equation}
\sigma_{4-1}/\sigma_c ={15\sqrt{3} \over 16}\left ((3\sqrt{3} - \sqrt{2})
({g^2 \over h^24} + 9/20) \right )^{-1} < 1
\end{equation}

 The fact that the steepest direction for $S = S_c$ is $G_{4-1}$ symmetric
is independent of the detailed
structure of the superpotential. To see this add to the superpotential
an arbitrary number of self-interaction terms. These are the all possible
polynomials made from independent holomorphyc invariants
$ A_n = {\rm Tr}\sigma^n$
\begin{equation}
W = {g\over 2}S{\rm Tr }\Sigma^2  - SM^2 +
 h_{n_1,..n_m}A_1^{n_1}...A_m^{n_m}
\end{equation}
The key point is that the lowest non-trivial invariant in the potential
is non-hermitian and has a form
\begin{equation}
 S^*\theta^n\phi^m\chi^l \sigma^*\sigma^k + h.c.
\end{equation}
where $n, m, k, l$ are integers and $\chi = {\rm Tr}\Sigma^5/\sigma^5$.
Thus, for the small values of $\sigma$ it dominates and its phase
automatically will be adjusted to negative. So the energetically most 
attractive
orientation will be the one that maximizes this term's absolute value. This is
the $4-1$ direction, for which all orbit parameters 
take larger values \footnote{If the dependence 
on $\chi$ is included, in principle, there can exist
other phenomenologically unacceptable local minima, e.g. with unbroken
$SU(2)\otimes SU(2)\otimes U(1)\otimes U(1)$.}.

Neglecting the possibility of `overshooting' in the $3-2-1$ vacuum, there thus
appears to be no away around this phenomenological problem in the
context of the standard field content and superpotential given above.
In order to avoid this phenomenological disaster then one is driven to two
possibilities.  Either the field  content of the theory must be made more
complicated, introducing extra SU(5) fields 
for example (see below), or the canonical normalization 
of the kinetic term in the Lagrangian must be altered, as one might
expect would result if some strongly coupled sector is responsible for the
inflationary potential \cite{savas}.

Ending up in the wrong vacuum is a severe enough problem.
However, it is clear from the scenario described above, which we
reiterate is the standard hybrid inflation scenario, that the GUT phase 
transition
occurs 
{\it after} inflation has ended.  This of course implies that one is left
with the standard GUT monopole problem.

In order to resolve the monopole problem in the context of GUT hybrid 
inflation, we can think of two alternatives.\footnote{Once again we want to 
stress that this discussion is essential for the $SU(5)$ GUT, in which
there is only one GUT phase transition. In extended
GUTs such as $SO(10)$\cite{jeannerot} or $SU(6)$\cite{dvaliriotto},
which assume more stages of symmetry breaking, 
it is possible to separate the inflationary
and GUT phase transitions.}
First, the $S$ field need not be an SU(5) singlet.  
In this case, SU(5) is broken before inflation, and one can imagine producing
monopoles before inflation, which then get 
inflated away as in the standard picture. 
An example of a superpotential which exhibits this behavior is:
\begin{equation}
W = S(g{\rm Tr}\Sigma^2/2 - M^2) + h {\rm Tr}\Sigma'\Sigma^2 + W_1(\Sigma'\Sigma)
\end{equation}
where $\Sigma'$ is another adjoint and $W_1$ is an arbitrary gauge invariant
superpotential such that 1) it allows for the $3-2-1$ vacuum; and 2)
$\partial_{\Sigma'}W_1 = \partial_{\Sigma}W_1 = 0$ for $\Sigma = 0$.
In this theory hybrid inflation 
can be driven by $\Sigma' = (2,2,2,-3,-3)\sigma'$, 
since for $\sigma' \pm S >> M$, $\Sigma$ will vanish
and the tree-level potential for $\sigma'$ and $S$ is essentially flat. Thus some combination
of these fields can play the role of the inflaton field.

Another alternative, which to our knowledge has not been discussed in the
literature before, is to suppose that an Affleck-Dine type mechanism
\cite{ad}
is exploited, even when the inflaton field $S$ is a singlet and 
the GUT transition 
happens after the inflation, 
or {\it even if inflation never happens}. 
This mechanism, was
originaly invented for generating the baryon asymmetry, 
and is based on the assumption that
some of the squark and slepton VEVs that parameterize the flat directions
of the SUSY vacua, are left far away from the origin during inflation and
later
perform coherent oscillations about it, generating 
the desired baryon to entropy
ratio. Such initial conditions are 
probable in supergravity theories, since the flat direction 
fields usually get large curvature (of order the Hubble
parameter) during inflation\cite{ddrt}, which may have either sign
depending on the precise structure of the K\"ahler potential.
We propose that this same
mechanism can cure the monopole problem, since the flat direction fields
can (generically do) break electromagnetism \cite{dvalkrau}.
Therefore the monopoles may never form. 
As a simple example (for further details see \cite{dvalkrau} 
consider the flat direction
parameterized by the invariant $\Phi^3 = 10^a\bar 5^c\bar 5^b$ 
where $10^a,~\bar 5^c,~\bar 5^b$ are matter superfields and $a,b,c$ are 
generation indexes (obviously $b \not= c$ due to the antisymmetric properties).
In the component form this flat direction reads
$10^a_{ik}=\left (\delta_{i1}\delta_{k2} - \delta_{i2}\delta_{k1} \right )
\Phi,~~~\bar 5^{ci} = \delta_{i1}\Phi,~~~\bar 5^{bk} =\delta_{k2}\Phi$
and breaks $SU(5)$ to $SU(3)_c$ during inflation. Thus, the phase transition 
at the end of inflation will not lead to the monopole formation.
It is worth noting that the problem can be solved even if the $\Phi$ direction
gets a nonzero VEV only after the GUT phase transition, e. g. as a result
of energy transfer from the oscillating inflaton field.
 During this time,
the U(1) breaking will confine monopoles, causing them to
eventually efficiently annihilate.
(Note that the time during which U(1)
must be broken must be fairly long if the reheating temperature
is sufficiently below the GUT scale \cite{gatkratern}.)
It is clear that this manifestation of
this mechanism (i.e. the $\Phi$ direction 
getting a nonzero VEV after the GUT transition)
can resolve the monopole problem even
if the GUT transition does not involve inflation.

As we have shown, while simple hybrid inflation models are well motivated in
principle on particle physics grounds and provide natural
inflation models with little or no fine tuning, the simplest GUT
prototype
models are in general phenomenological disasters.
In order to avoid these difficulties, significant
additional complexity must be introduced into these models,
or the inflationary phase transition and the GUT transition
must be decoupled. (In this regard we note that in 
order to produce acceptable density perturbations, the
GUT scale in even these hybrid models comes out {\it close to}, but
not exactly the standard SUSY SU(5) GUT scale. 
Some extra field content, or threshold corrections would
then be needed to truly tie the GUT and inflationary scales together.) 
Finally, our considerations have led us to recognize a new possible
solution to the monopole problem, either in the context of a hybrid GUT
inflation model, or even in a model with no inflation.  This mechanism
may thus be of interest well beyond 
the issue of the viability of hybrid inflation



\begin{thebibliography}{99}
\bibitem{inflation} A. H. Guth, Phys. Rev. D {\bf 23}, 347 (1981).
For a review of inflation and list of references, see A. D. Linde, {\it Particle Physics and Inflationary Cosmology} (Harwood Academic, New York, 1990); E.W.Kolb and M.S.Turner, {\it The Early
Universe} (Addison-Wesley, Reading, MA, 1990);

\bibitem{hybrid} A.D.Linde, Phys.Lett. {\bf B259} (1991) 38;
Phys. Rev. {\bf D49} (1994) 748.

\bibitem{copeland} E.J.Copeland, A.R.Liddle, D.H.Lyth,
E.D.Stewart and D.Wands, Phys.Rev. {\bf D49} (1994) 6410.

\bibitem{dss} G. Dvali, Q. Shafi and R. Schaefer,
{\it Phys.Rev.Lett.} {\bf 73} (1994) 1886;  G. Dvali, {\it  
Phys.Lett.} {\bf B387} (1996) 471.

\bibitem{lss} G. Lazarides, R.K. Schaefer and Q. Shafi, hep-ph/9608256.
\bibitem{weinberg} S. Weinberg, Phys. Rev. Lett. {\bf 48} (1982) 1776.
\bibitem{cl} S. Coleman and F. de Luccia, Phys. Rev. {\bf D21} (1980) 3305.
\bibitem{orbit} J. S. Kim, Nucl.Phys. {\bf B196} (1982) 285.

\bibitem{minima} See, e. g., L.-F. Li, Phys.Rev. {\bf D9}(1974) 1723;
M. Magg and Q. Shafi, Z.Phys. {\bf C4} (1980) 63;
J. S. Kim, Nucl.Phys. {\bf B197} (1982) 174.

\bibitem{gr} C. Panagiotakopoulos, hep-ph/9702433.

\bibitem{linderiotto} A. Linde and A. Riotto, hep-ph/9703209.

\bibitem{savas} S. Dimopoulos, G. Dvali and R. Rattazzi, preprint
hep-ph/9705348.

\bibitem{jeannerot} R. Jeannerot, Phys. Rev. {\bf D53} (1996) 5426.

\bibitem{dvaliriotto} G. Dvali and A. Riotto  hep-ph/9706408.

\bibitem{ad} I. Affleck and M. Dine, Nucl. Phys. {\bf B249} (1985) 361.

\bibitem{ddrt} G. Dvali, Preprint  hep-ph/9503259;
Phys. Lett. {\bf B355} (1995) 78; M. Dine, L. Randall and S. Thomas,
Phys. Rev. Lett. {\bf B75} (1995) 398.

\bibitem{dvalkrau} G. Dvali, L.M. Krauss, in preparation


\bibitem{gatkratern} E. Gates, L.M. Krauss, and J. Terning,
Phys. Lett. {\bf B28} (1992) 309


\end{thebibliography}
\end{document}